\documentclass[10pt]{article}
\usepackage{epsfig,amsmath,amsthm,graphicx,amssymb,overpic}
\usepackage{epsfig,amsmath,graphicx,amssymb,overpic,cite}
\usepackage{listings,diagbox}

\def\be{\begin{equation}}
\def\ee{\end{equation}}
\def\bee{\begin{eqnarray}}
\def\ene{\end{eqnarray}}
\def\bes{\begin{subequations}}
\def\ees{\end{subequations}}

\def\v{\vspace{0.1in}}

\newcommand{\PT}{\mathcal{PT}}

\begin{document}

%%%%%%%%%%%%%%%%%%%%%%%%%%%%%%%%%%%%%%%%%%%%%%%%%%%%%%%%%%%%%%%%%%%%%%%%%%%%%%%%%
\baselineskip=13pt
\renewcommand {\thefootnote}{\dag}
\renewcommand {\thefootnote}{\ddag}
\renewcommand {\thefootnote}{ }

\pagestyle{plain}

\begin{center}
\baselineskip=16pt \leftline{} \vspace{-.3in} {\Large \bf
 Deep learning neural networks for the third-order nonlinear Schr\"odinger equation: Solitons, breathers, and rogue waves} \\[0.2in]
\end{center}

\begin{center}
Zijian Zhou and  Zhenya Yan$^{*}$\footnote{$^{*}$Corresponding author. {\it Email address}: zyyan@mmrc.iss.ac.cn}  \\[0.1in]
{\it\small Key Laboratory of Mathematics Mechanization, Academy of Mathematics and Systems Science, \\ Chinese Academy of Sciences, Beijing 100190, China \\
 School of Mathematical Sciences, University of Chinese Academy of Sciences, Beijing 100049, China} \\
 %(Date:\,\, \today)
\end{center}

%\vspace{0.1in}

{\baselineskip=13pt

%\begin{tabular}{p{16cm}}
% \hline \\
%\end{tabular}

\vspace{0.18in}

%\begin{abstract} \small \baselineskip=12pt
\noindent {\bf Abstract}\, The third-order nonlinear Schr\"odinger equation (alias the Hirota equation) is investigated via deep leaning neural networks, which describes the strongly dispersive ion-acoustic wave in plasma and the wave propagation of ultrashort light pulses in optical fibers, as well as broader-banded waves on deep water. In this paper, we use the physics-informed neural networks (PINNs) deep learning method to explore the data-driven solutions (e.g., soliton, breather, and rogue waves) of the Hirota equation when the two types of the unperturbated and unperturbated (a $2\%$ noise) training data are considered. Moreover, we use the PINNs deep learning to study the data-driven discovery of parameters appearing in the Hirota equation with the aid of solitons.

%\vspace{0.1in} \noindent MSC: 37K15; 35Q53; 35Q15; 37K40

\vspace{0.1in} \noindent {\it Keywords:} Third-order nonlinear Schr\"odinger equation; deep learning; neural network; data-driven solutions

%\end{*abstract}

%\vspace{-0.05in}
%\begin{tabular}{p{16cm}}
%  \hline \\
%\end{tabular}

\section{Introduction}

As a fundamental and prototypical physical model, the one-dimensional cubic nonlinear Schr\"odinger (NLS) equation in the dimensionless form is
\bee\label{nls}
 iq_t+q_{xx}+2|q|^2q=0,\quad (x,t)\in \mathbb{R}^2,
\ene
where $q=q(x,t)$ denotes the complex field, and the subscripts stand for the partial derivatives with resect to the variables. Eq.~(\ref{nls}) can be used to describe the wave propagation in many fields of Kerr nonlinear and dispersion media such as plasmas physics, deep ocean, nonlinear optics, Bose-Einstein condensate, and even finance (see, e.g., Refs.~\cite{book0,nlsbook1,nls-ep1,nls-ep2,fnls1,fnls2,gp1,gp2,pr78,kelley65,dw68,mar75,book1,book2,book3,nls05,nls,yanfrw} and references therein). When the ultra-short laser pulse (e.g., 100 fs~\cite{book2}) propagation were considered, the study of the higher-order dispersive and nonlinear effects is of important significance, such as third-order dispersion, self-frequency shift, and self-steepening arising from the stimulated Raman scattering~\cite{hnls,hnls2,yan13}. The third-order NLS equation (alias the Hirota equation~\cite{hirota}) is also fundamental physical model. The Hirota equation and its extensions can also be used to describe the strongly dispersive ion-acoustic wave in plasma~\cite{gogoi10} and the  broader-banded waves on deep ocean~\cite{g-nls,g-nls-2}. The Hirota equation is completely integrable, and can be solved via the  bi-linear method~\cite{hirota}, inverse scattering transform~\cite{h-ist,yan2020}, and Darboux transform (see,
e.g., Refs.~\cite{h-breather,hrw1, hrw2, yan15, yanchaos15,yan2019}), and etc. Recently, we numerically studied the spectral signatures of the spatial Lax pair with distinct potentials (e.g., solitons, breathers, and rogue waves ) of the Hirota equation~\cite{yanchaos20}.

Up to now, artificial intelligence (AI) and machine learning (ML) have been widely used to powerfully deal with the big data, and
play an more and more important role in the various fields, such as language translation, computer vision,
speech recognition, and so on \cite{dl1, dl2}. More recently, the deep neural networks were presented to study the data-driven solutions and parameter discovery of nonlinear physical models~\cite{nn1,nn2,nn3,nn4,raiss18, siri2018, pang2019, zhang2019b,raiss19,lu2021,Han, raiss20, dl-3, yan20, yan21,dong19}. Particularly, the physics-informed neural networks (PINNs) technique~\cite{raiss18,raiss19,lu2021} were developed to study nonlinear partial differential equations.
In this paper, we would like to extend the PINNs deep learning method to investigate the data-driven solutions and parameter discovery for the focusing third-order nonlinear Schr\"odinger equation (alias the Hirota equation) with initial-boundary value conditions
\bee\label{hiro}
\left\{\begin{array}{l}
 iq_t+\alpha(q_{xx}+2|q|^2q)+i\beta(q_{xxx}+6|q|^2q_{x})=0,\quad x\in (-L, L),\quad t\in (t_0, T), \vspace{0.1in}\\
q(x,t_0)=q_0(x), \quad  x\in [-L, L], \vspace{0.1in}\\
q(-L, t)=q(L, t),\quad t\in [t_0, T],
\end{array}\right.
\ene
where $q=q(x,t)$ is a complex envelope field, $\alpha$ and $\beta$ are real constants for the second- and third-order dispersion coefficients, respectively. For $\beta=0$, the Hirota equation (\ref{hiro}) becomes a nonlinear Schr\"odinger (NLS) equation,
whereas $\alpha=0$, the Hirota equation (\ref{hiro}) reduces to the complex modified KdV equation~\cite{hirota}
\bee
 q_t+\beta(q_{xxx}+6|q|^2q_{x})=0.
\ene

The rest of this paper is arranged as follows. In Sec. 2, we simply introduce the PINN scheme, and apply it to investigate the data-driven soliton, breather, and rogue wave solutions of Eq.~(\ref{hiro}) with $\alpha=1,\, \beta=0.01$. In Sec. 3, we introduce the PINNs scheme, and apply it to study the data-driven parameter discovery of Eq.~(\ref{hiro}) with the aid of solitons. Finally,
we give some conclusions and discussions.

\section{The PINN scheme for the data-driven solutions}

\subsection{The PINNs scheme}

In this subsection, we would like to simply introduce the PINN deep learning method~\cite{raiss19} for the data-driven solutions. The main idea of the PINN deep learning method is to use a deep neural network to fit the solutions of Eq.~(\ref{hiro}). Let $q(x,t)=u(x,t)+iv(x,t)$ with $u(x,t),\,v(x,t)$ being its real and imaginary parts, respectively. The complex-valued PINN $F(x,t)=F_u(x,t)+iF_v(x,t)$ with $F_u(x,t),\,F_v(x,t)$ being its real and imaginary parts, respectively are written as
\bee
 \begin{array}{l}
 F(x,t):=iq_t+\alpha(q_{xx}+2|q|^2q)+i\beta(q_{xxx}+6|q|^2q_{x}), \v\\
 F_u(x,t):=-v_{t} + \alpha [u_{xx} +2(u^2+v^2)u]-\beta[v_{xxx} + 6(u^2 + v^2)v_{x}], \v\\
 F_v(x,t):= u_{t} +\alpha [v_{xx} + 2(u^2+v^2)v]+\beta[u_{xxx} + 6(u^2 + v^2)u_{x}],
 \end{array}
  \ene
and proceeded by approximating $q(x,t)$ by a complex-valued deep neural network. In the PINN scheme, the complex-valued neural network $q(x,t)=(u(x,t),\, v(x,t))$ can be written as
\begin{lstlisting}
def q(x, t):
    q = neural_net(tf.concat([x,t],1), weights, biases)
    u = q[:,0:1]
    v = q[:,1:2]
    return u, v
\end{lstlisting}

Based on the defined $q(x,t)$, the physics-informed neural network $F(x,t)$ can be taken as
\begin{lstlisting}
 def F(x, t):
     u, v = q(x, t)
     u_t = tf.gradients(u, t)[0]
     u_x = tf.gradients(u, x)[0]
     u_xx = tf.gradients(u_x, x)[0]
     u_xxx = tf.gradients(u_xx, x)[0]
     v_t = tf.gradients(v, t)[0]
     v_x = tf.gradients(v, x)[0]
     v_xx = tf.gradients(v_x, x)[0]
     v_xxx = tf.gradients(v_xx, x)[0]
     F_u = -v_t+alpha*(u_xx+2*(u**2+v**2)*u)-beta*(v_xxx+6*(u**2+v**2)*v_x)
     F_v = u_t+alpha*(v_xx+2*(u**2+v**2)*v)+beta*(u_xxx+6*(u**2+v**2)*u_x)
     return F_u, F_v
\end{lstlisting}

\begin{figure}[!t]
\hspace{2.5in}
\begin{center}
 {\scalebox{0.75}[0.75]{\includegraphics{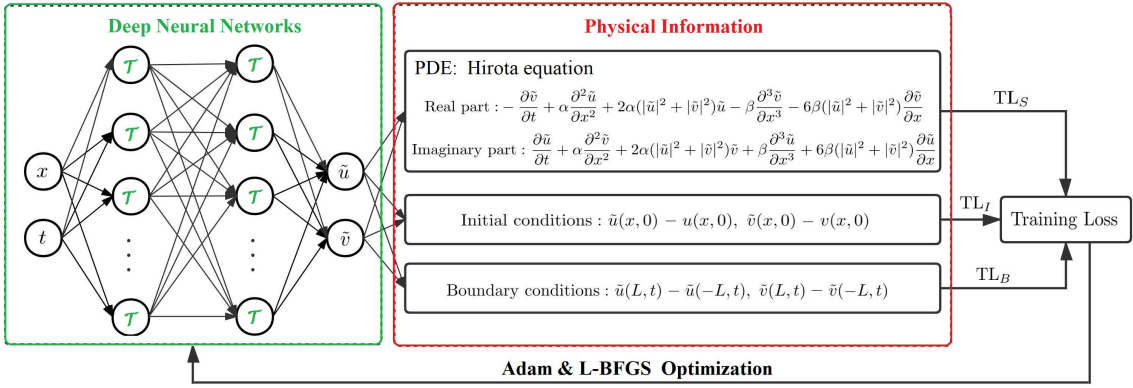}}}
\end{center}
\par
\vspace{0.05in}
\caption{\small The PINN scheme solving the Hirota equation (\ref{hiro}) with the initial and boundary conditions, where the activation function ${\cal T}=\tanh(\cdot)$.}
\label{fig-DNN}
\end{figure}

The shared parameters, weights and biases, between the neural network $\tilde{q}(x,t)=u(x,t) +iv(x,t)$ and $F(x,t)=F_u(x,t)+iF_v(x,t)$ can be learned by minimizing the whole training loss (TL), that is, the sum of the $\mathbb{L}^2$-norm training losses of the initial data (${\rm TL}_{I}$), boundary data (${\rm TL}_{B}$), and the whole equation $F(x,t)$ (${\rm TL}_{S}$)
 \bee
  {\rm TL}={\rm TL}_{I}+{\rm TL}_{B}+{\rm TL}_{S},
  \ene
where the mean squared (i.e., $\mathbb{L}^2$-norm) errors are chosen for them in the forms
\bee\begin{array}{l}
 {\rm TL}_{I}=\displaystyle\frac{1}{N_I}\sum_{j=1}^{N_I}\left(\left|u(x_{I}^j, t_0)-u_{0}^j\right|^2+\left|v(x_{I}^j, t_0)-v_{0}^j\right|^2\right),\v\\
 {\rm TL}_{B}=\displaystyle\frac{1}{N_B}\sum_{j=1}^{N_B}\left(\left|u(-L, t_{B}^j)-u(L, t_{B}^j)\right|^2+\left|v(-L, t_{B}^j)-v(L, t_{B}^j)\right|^2\right),\v\\
 {\rm TL}_{S}=\displaystyle\frac{1}{N_S}\sum_{j=1}^{N_S}\left(\left|F_u(x_S^j, t_S^j)\right|^2+\left|F_v(x_S^j, t_S^j)\right|^2\right)
\end{array}
\ene
with $\{x_I^j,\, u_{0}^j,\, v_{0}^j\}_{j=1}^{N_I}$ denoting the initial data ($q_0(x)=u_{0}(x)+iv_{0}(x)$),  $\{t_B^j,\, u(\pm L, t_{B}^j),\, v(\pm L, t_{B}^j\}_{j=1}^{N_B}$ standing for the periodic boundary data,  $\{x_S^j,\, t_S^j,\,F_u(x_S^j, t_S^j),\, F_v(x_S^j, t_S^j)\}_{j=1}^{N_S}$ representing the collocation points of $F(x,t)=F_u+iF_v$ within a spatio-temporal region
$(x,t)\in (-L, L)\times (t_0, T]$. All of these sampling points are generated using a space filling Latin Hypercube Sampling strategy~\cite{stein87}.

 %If $\alpha\neq1$, we use a simple transform $q(x,t)\rightarrow q(x,\tau)$, $\tau = \alpha t$ and $\beta\rightarrow\beta/\alpha$.
 We would like to discuss some data-driven solutions of Eq.~(\ref{hiro}) by the deep learning method.
 Here we choose a 5-layer deep neural network with 40 neurons per layer and a hyperbolic tangent activation function $\tanh(\cdot)$
 \bee
 \begin{array}{rl}
  A^{j+1}=&\tanh\left(W^{j+1}A^{j}+B^{j+1}\right)\v\\
    =&\displaystyle\left(\tanh\left(\sum_{s=1}^{m_{j}} w_{1s}^{j+1} a_s^{j}+b_1^{j+1}\right),\cdots,
    \tanh\left(\sum_{s=1}^{m_{j}} w_{m_{j+1}s}^{j+1} a_s^{j}+b_{m_j}^{j+1}\right)\right)^T,\quad j=0,1,2,..., M
    \end{array}
 \ene
 to approximate the learning solutions, where $A^j=(a_1^j, a_2^j,..., a_{m_j}^j)^T$ and $B^j=(b_1^j, b_2^j,..., b_{m_j}^j)^T$ denote the output and  bias column vectors of the $j$-th layer, respectively, $W^{j+1}=(w_{ks}^{j+1})_{m_{j+1}\times m_j}$ stands for the weight matrix of the $j$-th layer, $A^0=(x,t)^T$, $A^{M+1}=(u, v)^T$. The real and imaginary parts, $u(x,t)$ and $v(x,t)$,  of approximated solution $\tilde{q}(x,t)=u(x,t)+iv(x,t)$ are represented by the two outputs of one neural network (see Fig.~\ref{fig-DNN} for the PINN scheme).

In the following, we consider some fundamental solutions (e.g. soliton, breather, and rogue wave solutions) of Eq.~(\ref{hiro}) by using the PINNs deep leaning scheme. For the case $\alpha\beta\neq 0$ in Eq.~(\ref{hiro}), without loss of generality, we can take $\alpha=1,\, \beta=0.01$.

\subsection{The data-driven bright soliton}

The first example we would like to consider is the fundamental bright soliton of Eq.~(\ref{hiro})~\cite{hnls,hirota}
\begin{equation}\label{qbs}
    q_{bs}(x,t)={\rm sech}(x-\beta t)e^{it},
\end{equation}
where the third-order dispersion coefficient $\beta$ stands for the wave velocity, and the sign of $\beta$ represents the direction of wave propagation [{\it right-going} ({\it left-going}) travelling wave soliton for $\beta>0$ ($\beta<0)$].

We here choose $L=10,\, t_0=0,\, T=5$, and will consider this problem by choosing two distinct kinds of initial sample points: In the first case, we will choose the $N_{I}=100$ random sample points from the initial data $ q_{bs}(x,t=0)$ with $x\in [-10, 10]$. But in the second case, we only choose $N_{I}=5$ sample points from the initial data $ q_{bs}(x,t=0)$ with 5 equidistant and symmetric  points $x\in\{-5, -2.5, 0, 2.5, 5\}$. In the both cases, we use the same $N_{B}$=200 periodic boundary random sample points and $N_S=10,000$ random sample points in the solution region $\{(x,\,t,\, q_{bs}(x,t))|(x,t)\in [-10, 10]\times [0, 5]\}$. It is worth mentioning that the $N_S=10,000$ sample points are obtained via the Latin Hypercube Sampling strategy \cite{stein87}.

We emulate the first case of initial data by using 10,000 steps Adam and 10,000 steps L-BFGS optimizations such that  Figs.~\ref{fig1-soliton}(a1-a3) and (b1-b3) illustrate the learning results starting from the unperturbated and perturbated ($2\%$ noise)
training data, respectively. The relative $\mathbb{L}^2-$norm errors of $q(x,t)$, $u(x,t)$ and $v(x,t)$, respectively, are $9.3183\cdot 10^{-3}$, $5.3270\cdot 10^{-2}$, $3.8502\cdot 10^{-2}$ in Figs.~\ref{fig1-soliton}(a1-a2), and  $7.0707\cdot 10^{-3}$, $2.4057\cdot 10^{-2}$, $1.6464\cdot 10^{-2}$ in Figs.~\ref{fig1-soliton}(b1-a2).
Similarly,  we use the 20,000 steps Adam and 50,000 steps L-BFGS optimizations for the second case of initial data such that Figs.~\ref{fig1-soliton}(c1-c3) and (d1-b3) illustrate the learning results starting from the unperturbated and perturbated
training data, respectively. The relative $\mathbb{L}^2-$norm errors of $q(x,t)$, $u(x,t)$ and $v(x,t)$, respectively, are $1.8822\cdot 10^{-2}$, $4.9227\cdot 10^{-2}$, $4.0917\cdot 10^{-2}$ in Figs.~\ref{fig1-soliton}(c1-c2), and  $2.5427\cdot 10^{-2}$, $3.4825\cdot 10^{-2}$, $2.5983\cdot 10^{-2}$ in Figs.~\ref{fig1-soliton}(d1-d2).
Notice that those total learning times are (a) 717s, (b) 741s, (c) 1255s, and (d) 1334s, respectively, by using a Lenovo notebook with a 2.6GHz six-cores i7 processor and a RTX2060 graphics processor.

{\it Remark.}  In each step of the L-BFGS optimization, the program is stop at
\begin{equation}\label{stop}
    \frac{|{\rm loss}(n)-{\rm loss}(n-1)|}{{\rm max}(|{\rm loss}(n)|,|{\rm loss}(n-1)|,1)}<1.0\times {\rm np.finfo(float).eps},
\end{equation}
where the ${\rm loss}(n)$ represents the value of loss function in the $n$-th step L-BFGS optimization, and $1.0\times {\rm np.finfo(float).eps}$ represent Machine Epsilon. When the relative error between $loss(n)$ and $loss(n-1)$ less than Machine Epsilon, procedure would be stop. This is why the computation times are different for each test by using the same step optimization.

\begin{figure}[!t]
\begin{center}
\vspace{0.05in} %\hspace{-0.05in}
{\scalebox{0.6}[0.65]{\includegraphics{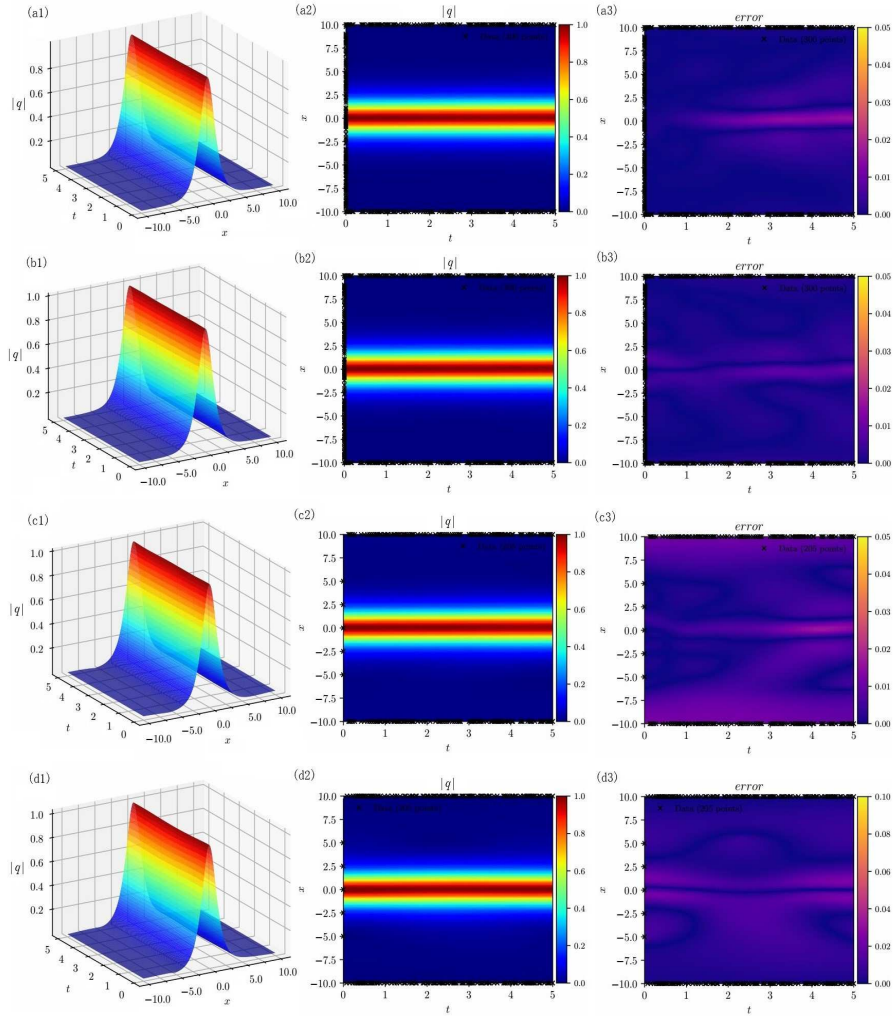}}}
\end{center}
\par
\vspace{-0.1in}
\caption{\small Data-driven soliton of the Hirota equation (\ref{hiro}): (a1,a2) and (b1,b2) the learning solutions
arising from the unpeturbated and perturbated ($2\%$) training data related to the first case of initial data, respectively;
(c1,c2) and (d1,d2) the learning solutions
arising from the unpeturbated and perturbated ($2\%$) training data related to the first case of initial data, respectively;
(a3, b3, c3, d3) the absolute values of the errors between the modules of exact and learning solutions. The relative $\mathbb{L}^2-$norm errors of $q(x,t)$, $u(x,t)$ and $v(x,t)$, respectively, are (a1-a3) $9.3183\cdot 10^{-3}$, $5.3270\cdot 10^{-2}$, $3.8502\cdot 10^{-2}$, (b1-b3) $7.0707\cdot 10^{-3}$, $2.4057\cdot 10^{-2}$, $1.6464\cdot 10^{-2}$, (c1-c3) $1.8822\cdot 10^{-2}$, $4.9227\cdot 10^{-2}$, $4.0917\cdot 10^{-2}$, (d1-d3) $2.5427\cdot 10^{-2}$, $3.4825\cdot 10^{-2}$, $2.5983\cdot 10^{-2}$.}
\label{fig1-soliton}
\end{figure}

\subsection{The data-deriven AKM breather solution}

The second example we would like to study is the AKM breather (spatio-temporal periodic pattern) of Eq.~(\ref{hiro})~\cite{h-breather}
\begin{equation}\label{qakm}
    q_{akm}(x,t)=\frac{\cosh(\omega t-2ic)-\cos(c)\cos(p\xi)}{\cosh(\omega t)-\cos(c)\cos(p\xi)}e^{2it},
\end{equation}
where $\xi=x-2\beta[2+\cos(2c)t]$, $\omega=2\sin(2c)$, $p=2\sin(c)$, and $c$ is a real constant. The wave velocity and wavenumber of this periodic wave are $2\beta (2+\cos(2c))$ and $p$, respectively. This AKM breather differs from the Akhmediev breather (spatial periodic pattern) of the NLS equation because Eq.~(\ref{hiro}) contains the third-order coefficient $\beta$. In this example, we assume $\beta = 0.01$ again. When $t\rightarrow \infty$, $|q_{akm}(x,t)|^2\rightarrow 1$. If $\beta\rightarrow 0$, we have $\xi\rightarrow x$, and then AKM breather almost becomes the Akhmediev breather.

We here choose $L=10$ and $t\in [-3, 3]$, and choose the $N_I=100$ random sample points from the initial data $ q_{akm}(x,t=0)$, $N_B=200$ random sample points from the periodic boundary data, and $N_S=10,000$ random sample points in the solution region $(x,t)\in [-10, 10]\times [-3, 3]$. We use the 20,000 Adam and 50,000 L-BFGS optimizations to learn the solutions from the unperturbated and perturbated  (a $2\%$ noise) initial data. As a result, Figs.~\ref{fig2-akm} (a1-a3) and (b1-b3) exhibit the leaning results for the unperturbated and perturbated  (a $2\%$ noise) cases, respectively.  The relative $\mathbb{L}^2-$norm errors of $q(x,t)$, $u(x,t)$ and $v(x,t)$, respectively, are (a) $1.1011\cdot 10^{-2}$, $3.5650\cdot 10^{-2}$, $5.0245\cdot 10^{-2}$, (b) $1.3458\cdot 10^{-2}$, $5.1326\cdot 10^{-2}$, $7.0242\cdot 10^{-2}$. The learning times are 2268s and 1848s, respectively.

\begin{figure}[!t]
\begin{center}
\vspace{0.05in} %\hspace{-0.05in}
{\scalebox{0.6}[0.6]{\includegraphics{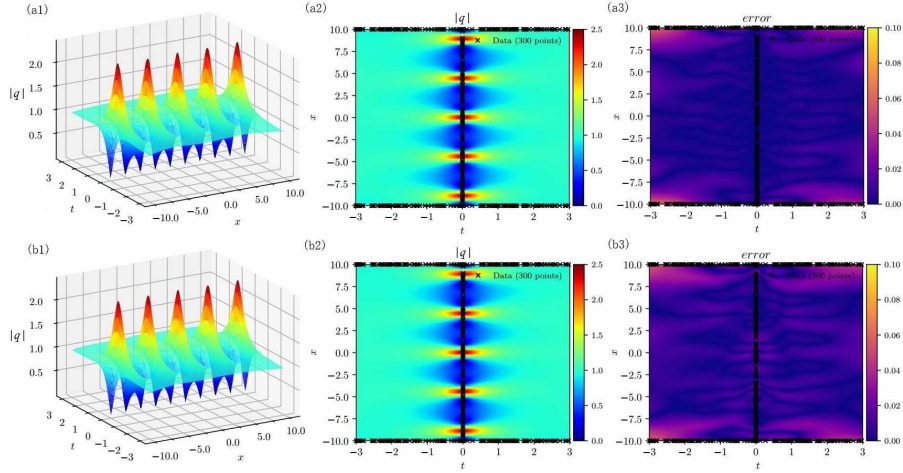}}}
\end{center}
\par
\vspace{-0.1in}
\caption{\small Learning breathers related to the AKM breather (\ref{qakm}) of the Hirota equation (\ref{hiro}). (a1-a3) the
unperturbbated case, (b1-b3) the $2\%$ perturbated case. The relative $\mathbb{L}^2-$norm errors of $q(x,t)$, $u(x,t)$ and $v(x,t)$, respectively, are (a1-a3) $1.1011\cdot 10^{-2}$, $3.5650\cdot 10^{-2}$, $5.0245\cdot 10^{-2}$, (b1-b3) $1.3458\cdot 10^{-2}$, $5.1326\cdot 10^{-2}$, $7.0242\cdot 10^{-2}$. }
\label{fig2-akm}
\end{figure}

\subsection{ The data-driven rogue wave solution}

The third example is a fundamental rogue wave solution of Eq.~(\ref{hiro}), which can be generated when one takes $c\rightarrow 0$ in the AKM breather (\ref{qakm}) in the form~\cite{hrw1}
\begin{equation}\label{qrw}
    q_{rw}(x,t)=\bigg[1-\frac{4(1+4it)}{4(x-6\beta t)^2+16t^2+1}\bigg]e^{2it}.
\end{equation}
As $|x|,|t|\to\infty$, $|q_{rw}|\to 1$, and ${\max}_{x,t}|q|=3$.

We here choose $L=2.5$ and $t\in [-0.5, 0.5]$, and consider $q_{rw}(x,t=-0.5)$ as the initial condition. We still choose $N_{I}=100$ random sample points from the initial data $q_{rw}(x,t=-0.5)$, $N_B=200$ random sample points from the periodic boundary data, and $N_S=10,000$ random sample points in the solution region $(x,t)\in [-2.5, 2.5]\times [-0.5, 0.5]$.
 We use the 20,000 steps Adam and 50,000 steps L-BFGS optimizations to learn the rogue wave solutions from the unperturbated and perturbated  (a $2\%$ noise) initial data, respectively. As a result, Figs.~\ref{fig3-rw}(a1-a3) and (b1-b3) exhibit the leaning results for the unperturbated and perturbated  (a $2\%$ noise) cases, respectively. The relative $\mathbb{L}^2-$norm errors of $q(x,t)$, $u(x,t)$ and $v(x,t)$, respectively, are (a) $6.7597\cdot 10^{-3}$, $8.8414\cdot 10^{-3}$, $1.6590\cdot 10^{-2}$, (b) $3.9537\cdot 10^{-3}$, $5.8719\cdot 10^{-3}$, $9.0493\cdot 10^{-3}$.  The learning times are 1524s and 1414s, respectively.

\begin{figure}[!t]
\begin{center}
\vspace{0.05in} %\hspace{-0.05in}
{\scalebox{0.6}[0.6]{\includegraphics{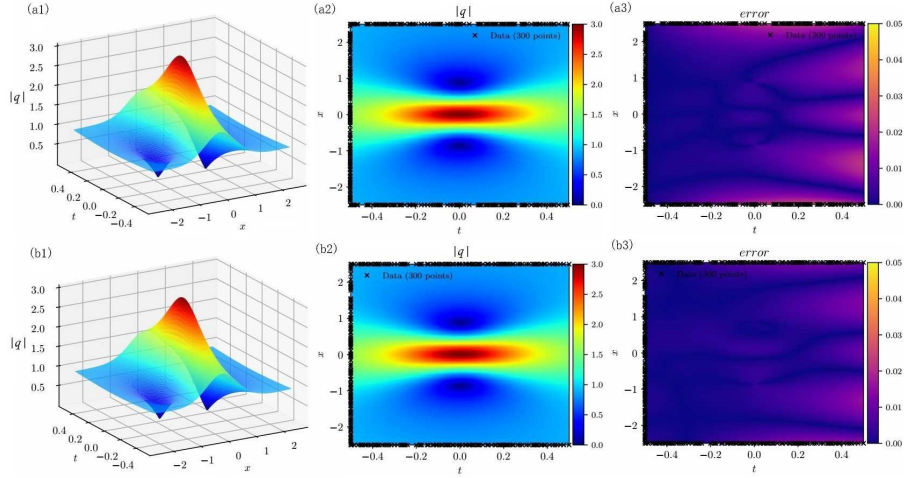}}}
\end{center}
\par
\vspace{-0.1in}
\caption{\protect\small Learning rogue wave solution related to Eq.~(\ref{qrw}) of the Hirota equation (\ref{hiro}).
(a1-a3) the unperturbated case, (b1-b3) the $2\%$ perturbated case. The relative $\mathbb{L}^2-$norm errors of $q(x,t)$, $u(x,t)$ and $v(x,t)$, respectively, are (a1-a3) $6.7597\cdot 10^{-3}$, $8.8414\cdot 10^{-3}$, $1.6590\cdot 10^{-2}$, (b1-b3) $3.9537\cdot 10^{-3}$, $5.8719\cdot 10^{-3}$, $9.0493\cdot 10^{-3}$. }
\label{fig3-rw}
\end{figure}

\section{The PINNs scheme for the data-driven parameter discovery}

In this section, we apply the PINNs deep learning method to study the data-driven parameter discovery of the Hirota equation (\ref{hiro}).
In the following, we use the deep learning method to identify the parameters $\alpha$ and $\beta$ in the Hirota equation (\ref{hiro}). Moreover, we also use this method to identify the parameters of the high-order terms of Eq.~(\ref{hiro}).

\subsection{The data-driven parameter discovery for $\alpha$ and $\beta$}

Here we would like to use the PINNs deep learning method to identify the coefficients $\alpha,\, \beta$ of second- and third-order dispersive terms in the Hirota equation
  \begin{equation}\label{hiro1}
    iq_t+\alpha(q_{xx}+2|q|^2q)+i\beta(q_{xxx}+6|q|^2q_{x})=0,
\end{equation}
 where $\alpha,\,\beta$ are the unknown real-valued parameters.

  Let $q(x,t)=u(x,t)+iv(x,t)$ with $u(x,t),\,v(x,t)$ being its real and imaginary parts, respectively, and  the PINNs $F(x,t)=F_u(x,t)+iF_v(x,t)$ with $F_u(x,t),\,F_v(x,t)$ being its real and imaginary parts, respectively, be
 \bee\label{f-eq}
 \begin{array}{l}
 F(x,t):= iq_t+\alpha(q_{xx}+2|q|^2q)+i\beta(q_{xxx}+6|q|^2q_{x}), \v\\
 F_u(x,t):= -v_{t} + \alpha [u_{xx} +2(u^2+v^2)u]-\beta[v_{xxx} + 6(u^2 + v^2)v_{x}], \v\\
 F_v(x,t):= u_{t} +\alpha [v_{xx} + 2(u^2+v^2)v]+\beta[u_{xxx} + 6(u^2 + v^2)u_{x}],
 \end{array}
  \ene
 Then the deep neural network is used to learn $\{u(x,t),\, v(x,t)\}$ and parameters $(\alpha,\, \beta)$ by minimizing the mean squared error loss
 \bee \label{2-loss}
 {\rm TL}={\rm TL}_q+{\rm TL}_p
 \ene
 with
\begin{equation} \label{2-lossg}
\begin{aligned}
    {\rm TL}_q&=\frac{1}{N_p}\sum_{j=1}^{N_p}\left(|u(x^j,t^j)-u^j|^2+|v(x^j,t^j)-v^j|^2\right),\\
    {\rm TL}_p &=\frac{1}{N_p}\sum_{j=1}^{N_p}\left(|F_u(x^j,t^j)|^2+|F_v(x^j,t^j)|^2\right),
    \end{aligned}
\end{equation}
where $\{x^j,\, t^j,\, u^j,\,v^j\}_{i=1}^{N_p}$ represents the training data on the real part and imaginary part of exact solution $u(x,t),\, v (x,t)$ given by Eq.~(\ref{qbs}) with $\alpha=1,\, \beta=0.5$ in $(x,t)\in [-8,8]\times[-3, 3]$, and $u(x^j,t^j),\, v(x^j,t^j)$ are real and imaginary parts of the approximate solution $q(x,t)=u(x,t)+iv(x,t)$.

To study the data-driven parameter discovery of the Hirota equation (\ref{hiro}) for $\alpha,\, \beta$,
we generate a training data-set by using the Latin Hypercube Sampling strategy to randomly select randomly choosing $N_q=10,000$ points in the solution region arising from
the exact bright soliton (\ref{qbs}) with $\alpha=1,\, \beta=0.5$ and $(x,t)\in [-8,8]\times[-3, 3]$.
Then the obtained data-set is applied to train an 8-layer deep neural network with 20 neurons per layer and a same hyperbolic tangent activation function to approximate
the parameters $\alpha,\, \beta$ in terms of minimizing the mean squared error loss given by Eqs.~(\ref{2-loss}) and (\ref{2-lossg}) starting from $\alpha=\beta=0$ in Eq.~(\ref{f-eq}). We here use the 20,000 steps Adam  and 50,000 steps L-BFGS optimizations.

Table~\ref{table1} illustrates the learning parameters $\alpha, \beta$ in Eq.~(\ref{hiro1}) under the cases of the data without perturbation and a $2\%$ perturbation, and their errors of $\alpha,\,\beta$ are 3.85$\times10^{-5}$, 7.48$\times10^{-5}$ and  3.31$\times10^{-4}$,  2.89$\times10^{-4}$, respectively. Fig.~\ref{fig4-ab} exhibits the learning  solutions and the relative $\mathbb{L}^2-$norm errors of $q(x,t)$, $u(x,t)$ and $v(x,t)$: (a1-a2) $7.0371\cdot 10^{-4}$, $1.0894\cdot 10^{-3}$, $1.0335\cdot 10^{-3}$; (b1-b2) $9.4420\cdot 10^{-4}$, $1.4055\cdot 10^{-3}$, $1.2136\cdot 10^{-3}$, where the training times are (a1-a2) 1510s and (b1-b2) 3572s, respectively.

\begin{figure}[!t]
\begin{center}
%\vspace{0.05in} %\hspace{-0.05in}
{\scalebox{0.56}[0.56]{\includegraphics{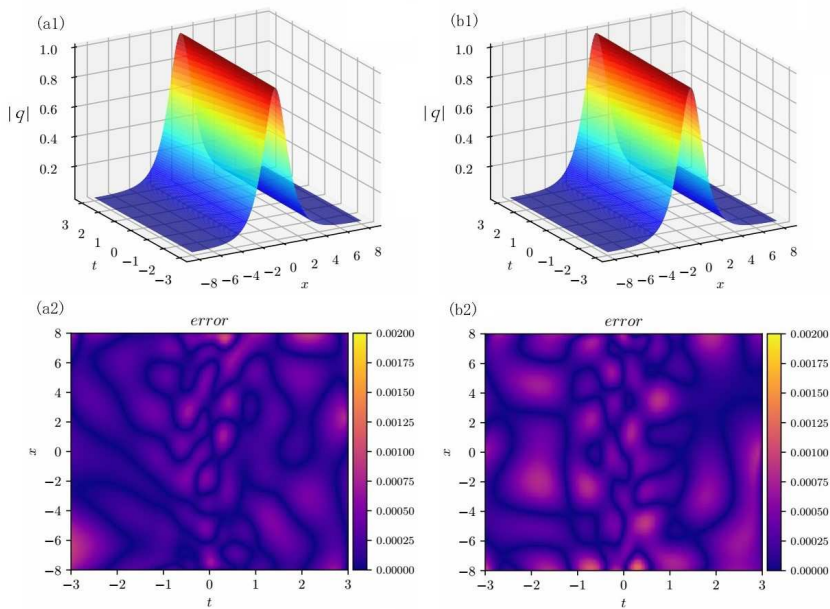}}}
\end{center}
\par
\vspace{-0.1in}
\caption{\protect\small Data-driven parameter discovery of $\alpha$ and $\beta$ in the sense of soliton (\ref{qbs}).
(a1-a2) soliton without perturbation. (b1-b2) soliton with a $2\%$ noise. (a2, b2) the absolute value of difference
between the modules of exact and learning solitons. The relative $\mathbb{L}^2-$norm errors of $q(x,t)$, $u(x,t)$ and $v(x,t)$, respectively, are (a1-a2) $7.0371\cdot 10^{-4}$, $1.0894\cdot 10^{-3}$, $1.0335\cdot 10^{-3}$, (b1-b2) $9.4420\cdot 10^{-4}$, $1.4055\cdot 10^{-3}$, $1.2136\cdot 10^{-3}$.}
\label{fig4-ab}
\end{figure}

\begin{table}
	\centering
    \setlength{\tabcolsep}{8pt}% column separation
    \renewcommand{\arraystretch}{1.4}%row space
	\caption{Comparisons of $\alpha$, $\beta$ and their errors in the different training data-set via deep learning. \vspace{0.05in}}
	\begin{tabular}{cccccc} \hline\hline
	 Case & Solution\qquad & $\alpha$ \qquad & error of $\alpha$ \qquad & $\beta$ \qquad & error of $\beta$ \\  \hline
  1& exact soliton \quad & 1 & 0 & 0.5 & 0 \\
  2 & soliton without perturbation \quad\quad& 1.00004 & 3.85$\times10^{-5}$ & 0.05008 & 7.48$\times10^{-5}$ \\
  3 &  soliton with a 2$\%$ perturbation \quad\quad & 0.99967 \qquad & 3.31$\times10^{-4}$ \qquad & 0.05029 \qquad & 2.89$\times10^{-4}$ \\
    \hline\hline
	\end{tabular}
	\label{table1}
\end{table}

\begin{figure}[!t]
\begin{center}
\vspace{0.05in} %\hspace{-0.05in}
{\scalebox{0.6}[0.6]{\includegraphics{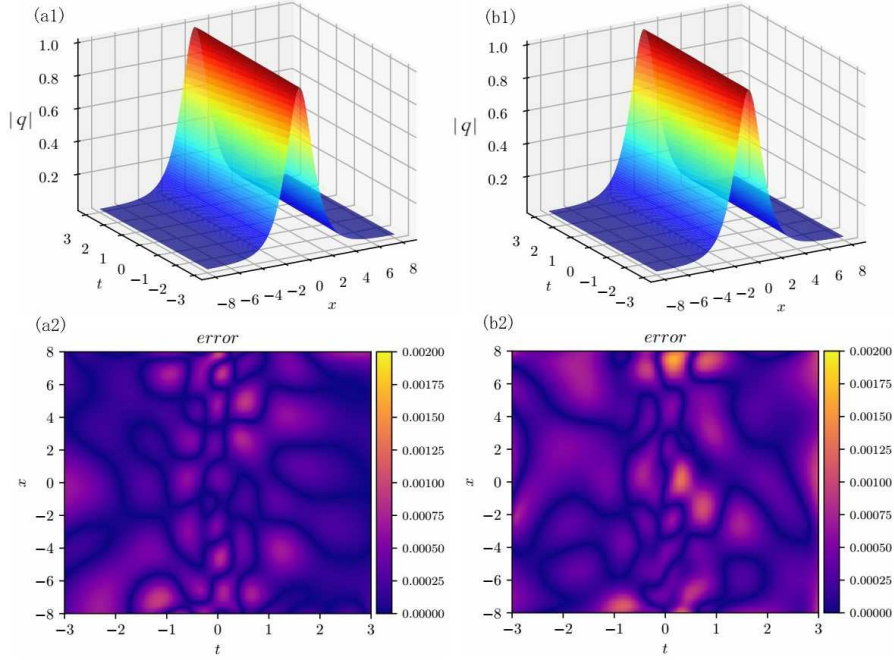}}}
\end{center}
\par
\vspace{-0.1in}
\caption{\protect\small Data-driven parameter discovery of $\mu$ and $\nu$ in the sense of soliton (\ref{qbs}). (a)(b) display the learning result under soliton data set. (a1-a2) are calculated without perturbation. (b1-b2) are calculated with 2$\%$ perturbation. (a2) and (b2) exhibit absolute value of difference between real solution and the function represented by the neural network. The relative $\mathbb{L}^2-$norm error of $q(x,t)$, $u(x,t)$ and $v(x,t)$, respectively, are (a1-a2) $8.0153\cdot 10^{-4}$, $1.0792\cdot 10^{-3}$, $1.2177\cdot 10^{-3}$, (b1-b2) $1.0770\cdot 10^{-3}$, $1.6541\cdot 10^{-3}$, $1.3370\cdot 10^{-3}$.}
\label{fig5-cd}
\end{figure}

\subsection{The data-driven parameter discovery for $\mu$ and $\nu$}

In what follows, we will study the learning coefficients of the high-order term in Eq.~(\ref{hiro}) via the deep learning method.
We consider the Hirota equation (\ref{hiro}) with two parameters in the form
\begin{equation}\label{hiro2}
    iq_t+q_{xx}+2|q|^2q+\frac{i}{2}(\mu q_{xxx}+\nu|q|^2q_{x})=0,
\end{equation}
where $\mu$ and $\nu$ are the unknown real constants of higher-order dispersion and nonlinear terms, respectively.

 Let $q(x,t)=u(x,t)+iv(x,t)$ with $u(x,t),\,v(x,t)$ being its real and imaginary parts, respectively, and  the PINNs $F(x,t)=F_u(x,t)+iF_v(x,t)$ with $F_u(x,t),\,F_v(x,t)$ being its real and imaginary parts, respectively, be
 \bee\label{f-eq}
 \begin{array}{l}
 F(x,t):= iq_t+q_{xx}+2|q|^2q+\frac{i}{2}(\mu q_{xxx}+\nu|q|^2q_{x}), \v\\
 F_u(x,t):= -v_{t} + u_{xx} +2(u^2+v^2)u-\frac12[\mu v_{xxx} + \nu(u^2 + v^2)v_{x}], \v\\
 F_v(x,t):= u_{t} +v_{xx} + 2(u^2+v^2)v+\frac12[\mu u_{xxx} + \nu(u^2 + v^2)u_{x}].
 \end{array}
  \ene
 Then the deep neural network is used to learn $\{u(x,t),\, v(x,t)\}$ and parameters $(\mu,\, \nu)$ by minimizing the mean squared error loss given by Eqs.~(\ref{2-loss}) and (\ref{2-lossg}).

 To illustrate the learning ability, we still use an 8-layer deep neural network with 20 neurons per layer. We choose $N_q=10,000$
sample points by the same way in the interior of solution region. The 20,000 steps Adam  and 50,000 steps L-BFGS optimizations are used in the training process. Table~\ref{table2} exhibits the training value and value errors of $\mu$ and $\nu$ in different training data set. And the results of neural network fitting exact solution are shown in Fig.~\ref{fig5-cd}. The training times are (a1-a2) 1971s and (b1-b2) 1990s, respectively.

\begin{table}[!t]
	\centering
    \setlength{\tabcolsep}{8pt}% column separation
    \renewcommand{\arraystretch}{1.4}%row space
	\caption{Comparisons of $\mu$, $\nu$ and their errors in the different training data-set via deep learning. \vspace{0.05in}}
	\begin{tabular}{cccccc} \hline\hline
	 Case & Solution\qquad & $\mu$ \qquad & error of $\mu$ \qquad & $\nu$ \qquad & error of $\nu$ \\  \hline
  1& exact soliton \quad & 1 & 0 & 1 & 0 \\
  2 & soliton without perturbation \quad\quad & 1.00370 & 3.69$\times10^{-3}$ & 6.03143 & 3.14$\times10^{-2}$ \\
  3 &  soliton with a 2$\%$ perturbation \quad\quad & 0.98159 & 1.84$\times10^{-2}$ & 5.88733 & 1.13$\times10^{-1}$ \\
    \hline\hline
	\end{tabular}
	\label{table2}
\end{table}

\section{Conclusions and discussions}

 In conclusion, we have explored the data-driven solutions and parameter discovery of the third-order nonlinear Schr\"odinger equation (alias the Hirota equation) via the deep learning method. We use the physics-informed neural networks (PINNs) deep learning method to study the data-driven fundamental solutions (e.g., soliton, breather, and rogue waves) of the Hirota equation, where the two types of the unperturbated and perturbated (a $2\%$ noise) training data are considered. Moreover, we use the PINNs deep learning to study the data-driven discovery of parameters appearing in the Hirota equation under the sense of its solitons.
 The PINN scheme can also be used to study the rogue waves of other nonlinear wave equations. \\

%\noindent \v {\bf Declaration of Competing Interest}

%All authors declare that they have no conflict of interest.  \\

\v \noindent {\bf Acknowledgments} \v

%The authors would like to thank the referees for the valuable suggestions and comments to improve the manuscript.
This work is supported by the NSFC under Grant Nos. 11925108 and 11731014.

\v \noindent {\bf Data availability statement} \v

The data that support the findings of this study are available
upon reasonable request from the authors.

\end{document}